\documentclass[twocolumn,trackchanges]{aastex62}

\usepackage{multirow}
\usepackage{textcomp}
\graphicspath{{./}{figures/}}

\shorttitle{Signature of a zero-metallicity hypernova}
\shortauthors{Sk\'{u}lad\'{o}ttir et al. 2021}

\begin{document}

\title{Zero-metallicity hypernova uncovered by an ultra metal-poor star\\ in the Sculptor dwarf spheroidal galaxy \footnote{Based on observations made with ESO VLT/X-SHOOTER at the La Silla Paranal observatory under program ID 0102.B-0786}}


\author{\'Asa Sk\'ulad\'ottir}
\affiliation{Dipartimento di Fisica e Astronomia, Universit\'{a} degli Studi di Firenze, Via G. Sansone 1, I-50019 Sesto Fiorentino, Italy\email{asa.skuladottir@unifi.it}}
\affiliation{INAF/Osservatorio Astrofisico di Arcetri, Largo E. Fermi 5, I-50125 Firenze, Italy}

\author{Stefania Salvadori}
\affiliation{Dipartimento di Fisica e Astronomia, Universit\'{a} degli Studi di Firenze, Via G. Sansone 1, I-50019 Sesto Fiorentino, Italy}
\affiliation{INAF/Osservatorio Astrofisico di Arcetri, Largo E. Fermi 5, I-50125 Firenze, Italy}

\author{Anish M. Amarsi}
\affiliation{Theoretical Astrophysics, Department of Physics and Astronomy, Uppsala University, Box 516, SE-751 20 Uppsala, Sweden}

\author{Eline Tolstoy}
\affiliation{Kapteyn Astronomical Institute, University of Groningen, PO Box 800, 9700AV Groningen, the Netherlands}

\author{Michael J. Irwin}
\affiliation{Institute of Astronomy, Madingley Road, Cambridge CB3 0HA, England}

\author{Vanessa Hill}
\affiliation{Laboratoire Lagrange, Universit\'{e} de Nice Sophia Antipolis, CNRS, Observatoire de la C\^{o}te d’Azur, CS34229, 06304 Nice Cedex 4, France}

\author{Pascale Jablonka}
\affiliation{Laboratoire d’astrophysique, Ecole Polytechnique Fédérale de Lausanne (EPFL), Observatoire, CH-1290 Versoix, Switzerland}
\affiliation{GEPI, Observatoire de Paris, CNRS, Université de Paris Diderot, F-92195 Meudon, Cedex, France}

\author{Giuseppina Battaglia}
\affiliation{Instituto de Astrof\'isica de Canarias, Calle Via L\'actea s/n, E-38206 La Laguna, Tenerife, Spain}
\affiliation{Universidad de La Laguna. Avda. Astrof\'isico Fco. S\'anchez, La Laguna, Tenerife, Spain}

\author{Else Starkenburg}
\affiliation{Kapteyn Astronomical Institute, University of Groningen, PO Box 800, 9700AV Groningen, the Netherlands}

\author{Davide Massari}
\affiliation{Kapteyn Astronomical Institute, University of Groningen, PO Box 800, 9700AV Groningen, the Netherlands}
\affiliation{INAF - Osservatorio di Astrofisica e Scienza dello Spazio di Bologna, Via Gobetti 93/3, I-40129 Bologna, Italy}

\author{Amina Helmi}
\affiliation{Kapteyn Astronomical Institute, University of Groningen, PO Box 800, 9700AV Groningen, the Netherlands}

\author{Lorenzo Posti}
\affiliation{Observatoire astronomique de Strasbourg, Université de Strasbourg, 11 rue de l’Université, 67000 Strasbourg, France}

\begin{abstract}
    Although true metal-free ``Population~III'' stars have so-far escaped discovery, their nature, and that of their supernovae,
    is revealed in the chemical products left behind in the next generations of stars.
    Here we report the detection of an ultra-metal poor star in the Sculptor dwarf spheroidal galaxy, AS0039.
    With [Fe/H]$_{\rm LTE}=-4.11$, it is the most metal-poor star so far discovered in any external galaxy.
    Contrary to the majority of Milky Way stars at this metallicity, AS0039 is clearly not enhanced in carbon, with [C/Fe]$_{\rm LTE}=-0.75$, and A(C)=+3.60, making it the lowest detected carbon abundance in any star to date.
    It furthermore lacks $\alpha$-element uniformity, having
    extremely low [Mg/Ca]$_{\rm NLTE}=-0.60$ and [Mg/Ti]$_{\rm NLTE}=-0.86$, 
    in stark contrast with the near solar ratios observed in C-normal stars within the Milky Way halo.
    The unique abundance pattern indicates that AS0039 formed out of material that
    was predominantly enriched by a $\sim$20$\,M_\odot$ progenitor star with 
    an unusually high explosion energy $E=10\times10^{51}$\,erg. 
    The star AS0039 is thus one of the first observational evidence for 
    zero-metallicity hypernovae and provides a unique opportunity to investigate the 
    diverse nature of Population~III stars.

\end{abstract}

\keywords{Sculptor dwarf elliptical galaxy (1436) --- Stellar abundances (1577) --- Population II stars (1284) --- Population III stars (1285) --- Hypernovae (775)}

\clearpage
\pagebreak

\section{Introduction} \label{sec:intro}

Great advances have been made in the discovery of ancient extremely metal-poor
stars ($\rm [Fe/H]<-3$) over the last decade. Within the Milky Way, there are now more than
a dozen stars known with $\rm [Fe/H]<-4.5$ \citep{Norris19}.
In dwarf spheroidal (dSph) and the ultra faint dwarf (UFD)
satellite galaxies, dozens of stars have now been discovered with 
$\rm [Fe/H]<-3$, with the hitherto most metal-poor star reaching $\rm [Fe/H]=-3.92\pm0.06$ \citep{Tafelmeyer10}.
Of these, almost half of the most metal-poor stars ($\rm [Fe/H]\lesssim-3.4$) belong
to the ancient Sculptor dSph galaxy, which is intrinsically metal-poor, 
$\text{\textlangle[Fe/H]\textrangle}=-1.8$
and is dominated by an old stellar population $>$10~Gyr
old \citep{Bettinelli19}.

Extremely metal-poor stars in the Milky Way
and its satellite dwarf galaxies show a dichotomy of abundance patterns, namely 
carbon-enhanced metal-poor (CEMP-no) stars 
($\rm [C/Fe]_{\rm LTE}>+0.7$, $\rm [Ba/Fe]_{\rm LTE}\leq0.0$),
and carbon-normal stars ($\rm [C/Fe]_{\rm LTE}\leq+0.7$).
The fraction of CEMP-no stars increases towards lower metallicity and reaches $\sim$70\% at $\rm[Fe/H]<-4$ \citep{Lee13,Yoon18}. The abundance patterns of CEMP-no stars have been successfully reproduced by theoretical yields of individual faint supernovae with
low explosion energy, along with mixing and fallback \citep{Iwamoto05}. These faint supernovae release significant amounts of the lighter elements, such as C, O, and Mg, but little amounts of heavier elements such as Fe. This results in high [C/Fe] ratios in their descendants. Often high 
values of other light-to-heavier element ratios such as [Mg/Fe] are also seen in CEMP-no stars and their $\alpha$-element abundant ratios are typically non-uniform  \citep{Norris13}.  

On the other hand, the C-normal stars show a remarkable uniformity
in the abundances of the $\alpha$- and the iron-peak
elements. This also holds for the most metal-poor C-normal stars found in the Milky Way halo,
below $\rm [Fe/H]<-4.5$ \citep{Caffau11,Starkenburg18}. The uniform abundance pattern indicates that these C-normal stars formed from a
well-mixed gas polluted by many supernovae \citep {Cayrel04}. 

Here, however, we report the discovery of an ultra metal-poor star in the Sculptor dSph galaxy, AS0039.  Having both low 
[C/Fe] and non-uniform $\alpha$-element abundance ratios, this star falls outside of the present dichotomy, thereby
indicating a different nucleosynthetic origin.\\

     \begin{figure}
   \centering
   \includegraphics[width=\hsize]{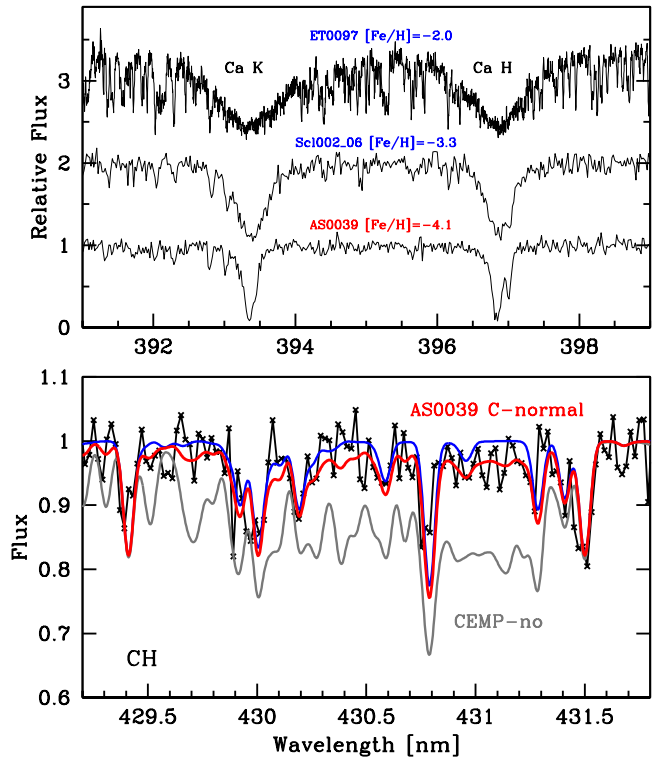} 
      \caption{The spectrum of AS0039. Top: Comparison to other metal-poor stars in Sculptor of similar temperature and gravity \citep{Starkenburg13,Skuladottir15}. 
      Bottom: Comparison of the CH G-band region of AS0039 to synthetic spectra: Red shows the best fit, blue is a synthetic spectrum with no CH lines, and gray represents a star born with the minimum amount of [C/Fe]=+0.7 to qualify as a CEMP-no star (accounting for mixing; \citealt{Placco14}).
      }
         \label{fig:spectra}
   \end{figure}

\section{Observations and data reduction} \label{sec:obs}
The star AS0039 was a part of a large survey (ESO ID 0102.B-0786) of the Ca~II
near-infrared triplet in Sculptor. 
This low-metallicty star was followed up with the ESO VLT/X-Shooter for a more detailed abundance analysis. 
The radial velocity of AS0039 from the X-Shooter spectrum was measured to be $v_{rad}=135\pm1$\,km/s. 
In addition to $v_{rad}$, the star's position, R.A.: 00:58:45.64, Dec.:$-$33:42:24.4; its location on a color magnitude
diagram as, $G=16.9$, $BP-RP=1.45$; and its proper motion,  
$\mu_{\alpha^*}=-0.01 \pm 0.06$\,mas/yr, $\mu_\delta=-0.16\pm0.05$\,mas/yr \citep{Gaia20},  all confirm AS0039 to be a member of the Sculptor
dwarf spheroidal galaxy. 
    
The X-Shooter spectra (see Fig.~\ref{fig:spectra})
were reduced with the most recent ESO pipeline. 
The present analysis focuses on the spectra from the UVB ($\lambda=300-560$\,nm, $R=5\,400$, $S/N=38$\,pix$^{-1}$ at 468\,nm) and the VIS ($\lambda=550-1030$\,nm, $R=8\,900$, $S/N=42$\,pix$^{-1}$ at
605\,nm) arms. Details of the observations and instrumental set-up are listed in the appendix.

\section{Spectral Analysis} \label{sec:abu}

\subsection{Atmospheric parameters and stellar models}
The effective temperature, $T_{\rm eff}=4377 \pm
81$\,K, of the star AS0039 was determined from {\it Gaia} eDR3
photometry \citep{Gaia20,Mucciarelli20}. The gravity, $\log g=0.8 \pm 0.1$,
was obtained through photometry, using the known distance to Sculptor (see \citealt{Skuladottir15}). Finally the
microturbulent velocity, $v_{mic}=2.0 \pm 0.1$\, km/s, was determined with an empirical
calibration \citep{Kirby09}. The best-fitting stellar atmosphere model was
taken from MARCS \citep{Gustafsson08} and all abundance analyses were
carried out using TURBOSPEC \citep{Plez12}, assuming local thermodynamic equilibrium (LTE),
as described in Sect.~\ref{sec:abuanalysis}.
These LTE abundances were then post-corrected, see Sect.~\ref{sec:nlte}.

 \begin{table}
\begin{center}
\caption{Chemical abundances of AS0039.
}
\label{tab:abu}
\begin{tabular}{lcrrr}
\hline
\noalign{\smallskip}
\multicolumn{1}{c}{Element} & \multicolumn{1}{c}{$\log\epsilon_{\odot}$} & \multicolumn{1}{c}{$\mathrm{[X/Fe]_{LTE}}$} & \multicolumn{1}{c}{$\mathrm{[X/Fe]_{NLTE}}$} & \multicolumn{1}{c}{$\sigma_{\text{NLTE}}$ [dex]} \\ 
\noalign{\smallskip}
\hline
\noalign{\smallskip}
Fe & $  7.46$ & $ -4.11$\tablenotemark{a} & $ -3.95$\tablenotemark{a} & $  0.17$ \\ 
\noalign{\smallskip}
\hline
\noalign{\smallskip}
C & $  8.46$ & $ -0.75$ & $ -0.91$ & $  0.22$ \\ 
Na & $  6.22$ & $  +0.02$ & $ -0.15$ & $  0.11$ \\ 
Mg & $  7.55$ & $ -0.09$ & $ -0.16$ & $  0.12$ \\ 
Si & $  7.51$ & $  +0.06$ & $ -0.15$ & $  0.28$ \\ 
Ca & $  6.30$ & $  +0.65$ & $  +0.44$ & $  0.11$ \\ 
Sc & $  3.14$ & $ -0.24$ & $ -0.40$ & $  0.28$ \\ 
Ti & $  4.97$ & $  +0.66$ & $  +0.70$ & $  0.13$ \\ 
Cr & $  5.62$ & $  +0.04$ & $  +0.58$ & $  0.18$ \\ 
Sr & $  2.83$ & $ -0.91$ & $ -1.07$ & $  0.22$ \\ 
\noalign{\smallskip}
\hline
\noalign{\smallskip}
Li & $  0.96$ & $ +3.65$\tablenotemark{b} & $ +3.52$\tablenotemark{b} & - \\ 
Al & $  6.43$ & $ -1.02$\tablenotemark{b} & $ -0.18$\tablenotemark{b} & - \\ 
Mn & $  5.42$ & $ -0.76$\tablenotemark{b} & $ +0.08$\tablenotemark{b} & - \\ 
Co & $  4.94$ & $ +0.07$\tablenotemark{b} & $ +0.91$\tablenotemark{b} & - \\ 
Ni & $  6.20$ & $ +0.21$\tablenotemark{b} & $ +0.21$\tablenotemark{b} & - \\ 
Ba & $  2.27$ & $ -1.36$\tablenotemark{b} & $ -1.52$\tablenotemark{b} & - \\ 
\noalign{\smallskip}
\hline
\noalign{\smallskip}
\end{tabular}
\end{center}
\tablenotetext{a}{results for $\mathrm{[Fe/H]}$ are shown} 
\tablenotetext{b}{upper limits are shown}
\end{table}

\subsection{LTE abundances}  \label{sec:abuanalysis}
All measured abundances are listed in Table~\ref{tab:abu} (see also Appendix~\ref{app:literature}). Not included are the
cases where only limits on the level of $\rm [X/Fe]>+1$ or higher were obtained.
The solar abundances were adopted from the recent compilation of \citet{Asplund21}, and all literature data discussed and shown in this paper has been put on the same scale.

The abundance of C was measured from the CH
molecular band at $430$\,nm. 
The strength of the CH molecular lines
can be affected by the assumed O abundance. Unfortunately, the \ion{O}{1} line at 
$630\,\mathrm{nm}$ only gave an upper limit of $\rm [O/Fe]<+2$.
Thus a value of 
$\rm [O/Fe]=+0.6$ was assumed for the synthetic spectra, since this corresponds to
the average value measured at $\rm [Fe/H]<-2$ in Sculptor \citep{Hill19}, and is
in good agreement with the Milky Way halo \citep{Amarsi19}.
The best fitting supernova model (Sect.~\ref{sec:formation}) 
implies $\rm [O/Fe]\approx+0.1$; nevertheless, 
a change of $\rm \Delta[O/Fe]=\pm0.5$\,dex fortunately only had a minor
effect, $\rm \Delta[C/Fe]=\mp0.04$\,dex, because the C/O ratio stays well
below unity in all cases. 
The N abundance could only be constrained through NH lines to $\rm [N/Fe]<+2.20$.

The Na abundance was measured from the \ion{Na}{1} D resonance lines at 589.0, and
589.6\,nm, which gave consistent results within error bars. For the odd elements
Al and K, only upper limits could be determined, $\rm [Al/Fe]<-1$ (\ion{Al}{1} 396.2\,nm), and $\rm [K/Fe]<+1.15$ (\ion{K}{1} 769.9\,nm). The abundance of Sc was
measured from two lines of \ion{Sc}{2}, at 424.7, and 431.4\,nm. The result between those two
lines differed by 0.4~dex, which is still consistent given the errors.

The abundances of three $\alpha$-elements were determined: Mg, Ca, and Ti.
Four \ion{Mg}{1} lines were measured, at 382.9, 383.8, 517.3, and 518.4\,nm.
The Ca abundance was determined
by two \ion{Ca}{1} lines (422.7, 616.2\,nm), and the near-infraread \ion{Ca}{2} triplet.
In total 15 \ion{Ti}{2} lines were used for the abundance determination of Ti.
A further three \ion{Ti}{1} lines were measured, but were ultimately dropped
due to large and uncertain non-LTE effects.

With the exception of Cr and Fe, only upper limits for
the iron-peak elements could be determined. Four \ion{Cr}{1} lines were used for the
determination of Cr, while 31 \ion{Fe}{1} lines were measured.  The upper limits for
Mn, Co, and Ni, based on lines of the neutral species,
are listed in Table~\ref{tab:abu}, but the upper limits for other elements (V,
Cu, Zn) were too high ($\rm [X/Fe]<+1.5$) to be
informative.

The only neutron-capture element that could be reliably
measured was Sr, which was based on two \ion{Sr}{2} lines (407.8, and 421.6\,nm) which
agreed within error bars. No Ba line was visible in the spectrum, but the \ion{Ba}{2}
line at 455.4\,nm\ gave the listed upper limit. Other upper limits of neutron-capture elements could not be determined to better than $\rm [X/Fe]<+1$, and were in most cases significantly higher.

The errors were determined as follows. For elements with
two or more atomic or ionic lines measured,
$N_l\geq2$, the error in $\log N_{X}/N_{H}$ was determined as the
standard error in the mean.  
For the remaining elements (C, Si), 
the error was determined based on the $\chi^2$ of the fits of the synthetic
spectra \citep{Skuladottir15b,Skuladottir17}.
The error in the adopted solar abundances \citep{Asplund21}
were added in quadrature to obtain errors in [X/H];
the error in [Fe/H] was similarly folded into the error estimate
for [X/Fe].
Random errors due to the stellar atmospheric parameters
were finally folded into these estimates; these were determined to be $\rm
\Delta[Fe/H]_{stellpar}=0.16$\,dex, and a typical value of $\rm
\Delta[X/Fe]_{stellpar}=0.06$\,dex was obtained for abundance ratios.

\subsection{Departure from LTE}    \label{sec:nlte}

The non-LTE corrections for Mg, Ca, and Fe were calculated specifically
    for this study, using the Balder code \citep{Amarsi18} (our custom version of Multi3D; \citealt{Leenaarts09}). 
    The general method of calculations follows that given in \citet{Amarsi20}; in particular,
    scattering by background lines, including Rayleigh scattering
    in the red wing of the Lyman series, was included as described therein.
    The calculations were performed on
    1D spherical MARCS model atmospheres
    of standard chemical composition,
    and used recent model atoms \citep{Asplund21}
    that adopt physically-motivated descriptions of inelastic collisions
    with neutral
    hydrogen.
    The equivalent widths were determined by direct integration,
    from which abundance corrections were obtained. These were 
    interpolated onto the stellar parameters
    $T_{\rm eff}=4377\,\mathrm{K}$ and $\log g=1.0$.
    
   The non-LTE corrections for the other measured elements were generally drawn
    from the literature: for Na from \citet{Lind11} via the INSPECT database\footnote{\url{http://www.inspect-stars.com}},
    for Si from \citet{Amarsi17}, and for Ti and Cr
    from \citet{Bergemann11} and \citet{Bergemann10} via the NLTE-MPIA database\footnote{\url{http://nlte.mpia.de}} \citep{Kovalev19}.
    The corrections for Sc and Sr were neglected
    for lack of a better alternative; in reality, slightly negative
    corrections might be expected for these low-excitation, ionised
    lines. We note that for Sc, rather
    minor corrections reaching down to $-0.04$ dex 
    have been noted in the literature \citep{Zhang14},
    while for Sr, data in the INSPECT database based on \citet{BergemannSr} extends
    down to $\log g=2.2$, where the mean correction is just $-0.03$ dex.

    For the elements with upper limits, only the non-LTE corrections
    for Li \citep{Wang21, Barklem21} were drawn from the literature.
    For Al, Mn, and Co, non-LTE corrections in the
    literature are large \citep{Nordlander17,BergemannMn,BergemannCo}, and rather uncertain
    owing to uncertainties in the atomic data and, in the case of Al, with the
    equivalent-width method; consequently, a conservative estimate of $+1$~dex
    was adopted here. For Ni, the mean abundance correction found for Fe was adopted
    ($+0.16$ dex). The correction for the ionised \ion{Ba}{2} resonance line was assumed to be
    negligible by virtue of the extremely low line strength.
    
Finally, we note that 3D effects are likely to bring the C abundance down \citep{Caffau11,Norris19}, and Fe slightly
up \citep{AmarsiFe}. Consequently, the 3D non-LTE [C/Fe] ratio is expected to be even lower than the 1D value listed in Table~\ref{tab:abu}.

\section{Results} \label{sec:res}
The spectrum of the newly discovered Sculptor star AS0039 (Fig.~\ref{fig:spectra})
reveals it to be an ultra metal-poor star, $\rm
[Fe/H]_{\rm LTE}=-4.11$ ($\rm[Fe/H]_{\rm NLTE}=-3.95$). The low [Fe/H] in combination with extremely low [C/H] and [Mg/H] (Table~\ref{tab:abu}) make AS0039 the most metal-poor star currently known in any external galaxy. 

While a large fraction ($\sim$70\%) of ultra metal-poor
stars \citep{Lee13,Yoon18} in the Milky Way are carbon-enhanced ($\rm [C/Fe]>+0.7$), this Sculptor star has
extremely low $\rm [C/Fe]_{\rm NLTE}=-0.91$ (Fig.~\ref{fig:spectra}). This corresponds to A(C)=+3.60, which is currently the lowest C~abundance detected in any galaxy, breaking the recent record, set by the Milky Way star  SPLUS J210428.01-004934.2, at A(C)=+4.34 \citep{Placco21}. The carbon abundance of AS0039 remains low even after correcting for mixing on the upper red giant branch (RGB),
which brings processed material to the surface, lowering the C and enhancing
the N abundances; following \citet{Placco14}, the initial birth composition
becomes $\rm [C/Fe]_{corr}=-0.29$. The low C in AS0039 is in line with other stars in
Sculptor, which seems to be relatively void of C-rich stars at low
metallicities, compared to the Milky Way \citep{Starkenburg13,Skuladottir15}.

        \begin{figure}
   \centering
   \includegraphics[width=\hsize]{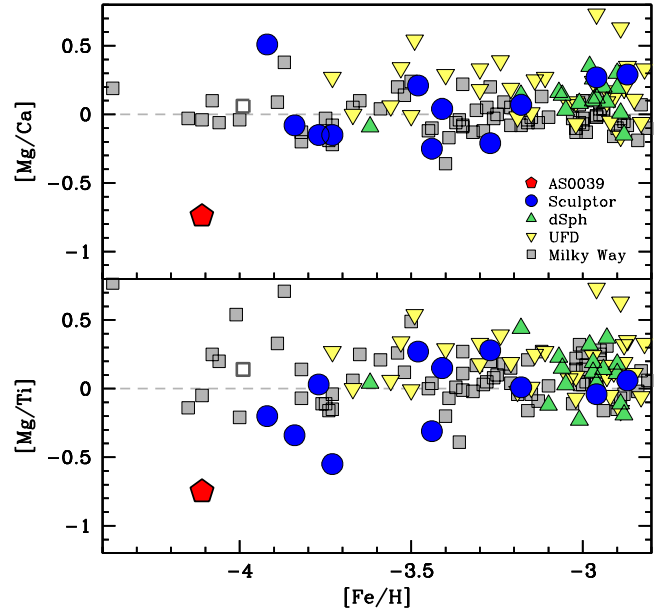} 
      \caption{Comparison of the 1D LTE $\alpha$-element ratios of AS0039 to those of the literature, [Mg/Ca] (top) and [Mg/Ti] (bottom).
		Classical 1D LTE abundances for RGB stars in Sculptor (blue), other dSph galaxies (green), the UFD galaxies (yellow), and C-normal RGB stars in the Milky Way \citep {Suda08}. The Milky Way star SPLUS J210428.01-004934.2 \citep{Placco21} is shown with a gray open square.
      }
         \label{fig:MgCaTi}
   \end{figure}

One of the most notable characteristics of the 
AS0039 abundance pattern is the low
[Mg/Fe]$_{\rm NLTE}=-0.16\pm0.12$, while C-normal, extremely metal-poor stars
in the Milky Way have a tight plateau \citep{Andrievsky10} of
super-solar [Mg/Fe]$_{\rm NLTE}=+0.57\pm0.13$. In the rare cases where Mg-poor
stars have been discovered, their depletion of Mg is always followed with low
abundances of other $\alpha$-elements, such as Ca and Ti. This is not the case
for AS0039, where both the [Ca/Fe] and [Ti/Fe] are higher than what is typical
for stars at these low metallicities, both in Sculptor and in the Milky
Way \citep{Cayrel04,Jablonka15}.

The $\alpha$-element ratios [Mg/Ca] and [Mg/Ti] for AS0039 are compared to the
published literature LTE values in Fig.~\ref{fig:MgCaTi}. The stars of the Milky
Way \citep{Cayrel04} sit at around $\rm[Mg/Ca]_{LTE}=-0.03\pm0.10$, which is in sharp
contrast with the low value of AS0039, $\rm[Mg/Ca]_{LTE}=-0.60\pm0.15$.
Abundance measurements in other stars in dwarf galaxies, both dSph and UFD,
also have a plateau of solar, or even slightly super-solar values of $\rm
[Mg/Ca]_{LTE}$. Similarly, AS0039 is a clear outlier from the normal trend with
$\rm[Mg/Ti]_{LTE}=-0.82\pm0.16$, while the Milky Way has an average value \citep{Cayrel04} of
$\rm[Mg/Ti]_{LTE}=+0.08\pm0.12$. The ratios of [Mg/Ti] in the Sculptor dSph
differ from other galaxies, both the Milky Way and other dwarf galaxies, with a
declining trend towards lower [Fe/H]. Combined with the lack of CEMP stars
in this galaxy, this is further evidence that the earliest chemical enrichment
of Sculptor was significantly different from that of the Milky Way. The lack of
known stars in other dwarf galaxies at $\rm [Fe/H]<-3.5$ makes it currently
impossible to conclude whether this trend of [Mg/Ti] is unique for Sculptor, or
a common trait of dSph, or even UFD galaxies.  To ensure a fair comparison, Fig.~\ref{fig:MgCaTi} is limited to RGB stars, but we note that the uniformity of $\alpha$-elements in the Milky Way C-normal stars extends down to the lowest $\rm[Fe/H]<-4.5$ \citep{Caffau11,Starkenburg18}.

The extreme $\alpha$-element ratios are not the only distinctive feature in the
abundance pattern of AS0039, which also has high $\rm [Cr/Fe]_{\rm
NLTE}=+0.58\pm0.18$ compared to the tight plateau around the solar ratio that
is observed in the metal-poor Milky Way halo \citep{Bergemann10}, with $\rm [Cr/Fe]_{
    \rm NLTE}=0.00\pm0.06$. Quite low upper limits were obtained for the odd
elements Al, Mn and Co, comparable to or lower than
the typically measured values in
Sculptor, while the odd element Na in AS0039 has a normal abundance for
extremely metal-poor stars. Finally, the extremely low abundance of the
neutron-capture elements, $\rm [Ba/H]\leq-5.5$ and $\rm [Sr/H]=-5.02\pm0.22$ are
typical for pristine second-generation stars \citep{Jablonka15}. 

On the whole, the abundance
pattern of AS0039 is starkly different
from all other extremely metal-poor stars
that have been discovered in the Milky Way or its satellite galaxies. 
This suggest that AS0039 was not formed from well-mixed material enriched by many supernovae,
instead it is showing a dominant contribution from only one single event.
This is in line with theoretical studies that have shown that a single supernovae is sufficient to reach metallicities of $\rm [Fe/H]=-4$ in dwarf galaxies \citep{Cooke14,Rossi21}. AS0039 therefore opens up a new window into the diverse properties of the Population~III stars.

 \begin{figure*}
   \centering
   \includegraphics[width=13cm]{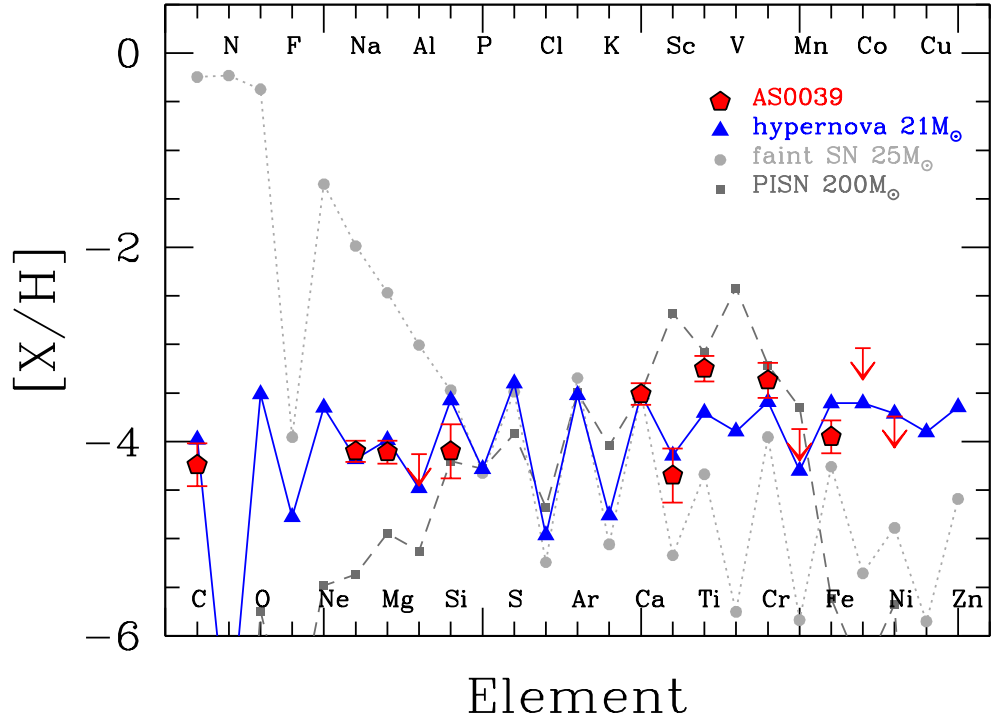} \caption{
      The NLTE abundance pattern of AS0039 (red pentagons) compared to theoretical models. Blue triangles show the best fitting model of a zero-metallicity 21\,M$_\odot$ hypernovae, of energy $10\times 10^{51}$\,erg. Also depicted are models for a PopIII faint supernova of  25\,M$_\odot$ with $E=0.7\times 10^{51}$\,erg (light grey circles), and pair instability supernova of 200\,M$_\odot$ (dark grey squares). All models are normalized to the measured [Ca/H]$_{\rm NLTE}$ of AS0039. The C abundance of AS0039 has been corrected for internal mixing \citep{Placco14}.
      }
    \label{fig:HN}
\end{figure*}

\section{Formation scenario} \label{sec:formation}

To investigate the formation scenario of AS0039, its observed abundances were
compared to nucleosynthetic stellar yields of 
Population~III supernovae over a large range of progenitor masses ($10-100$\,M$_\odot$),
explosion energies ($0.3-10\times10^{51}$\,erg), and internal
mixing efficiencies \citep{HegerWoosley10}, 
using the Starfit tool\footnote{\url{http://starfit.org}}.
The best fit revealed that the unique abundance pattern of AS0039 shows a
dominant signature of a $M=21$\,M$_\odot$ hypernovae, with $E=10\times
10^{51}$\,erg, see Fig.~\ref{fig:HN}. For confirmation, we also used our own fitting routine \citep{Salvadori19}, which gave the best fit of a hypernovae with $M=20\pm2$\,M$_\odot$, in good agreement with the result obtained with Starfit.

For comparison, we also show an example of a
model of a typical faint supernova (Fig.~\ref{fig:HN}), which have been very successful in
explaining the origin of CEMP-no stars \citep{Iwamoto05};
and of a pair-instability supernovae (PISN) of a typical mass $M=200$\,M$_\odot$. Such PISN enrich their surroundings with large amounts of Fe, and
their descendants are thus typically expected at much higher
metallicities, $\rm [Fe/H]\approx-2$ \citep{Salvadori19}. Both faint SNe and PISN can be excluded as having a dominant contribution to AS0039.

The high explosion energies of hypernovae permit simultaneously 
low [C/Fe], [Mg/Ca], and [Mg/Ti].  While C and Mg have hydrostatic origin, and are
mainly created in fusion throughout the lifetime of the star, both Ca and Ti
are so-called explosive $\alpha$-elements that mainly form in the supernova itself.
With the higher explosion energy of the
hypernovae, relatively high amounts of Ca and Ti are formed, resulting in 
a low [Mg/Ca,Ti] ratio, as is observed in AS0039. The opposite is true in faint SN, and close to half of CEMP-no stars have high $\rm [Mg/Ca]\gtrsim+0.5$ \citep{Norris13}. During these faint SN, only the outer C-rich layers are expelled, while the Fe-rich center of the star falls into the black hole, resulting in very high [C/Fe] ratios \citep{Iwamoto05}.
We note that none of the
available models \citep{HegerWoosley10} were able to reproduce the high [Ti/Fe] observed in
AS0039. However this is in line with other results, where theoretical models
are known to systematically under-produce Ti relative to
observations at all metallicities \citep{HegerWoosley10,Kobayashi20}.

\section{Discussion} \label{sec:dis}

With a striking signature of 
simultaneously low values of [C/Fe] and [Mg/Ca,Ti],  
the star AS0039 is thus the first unambiguous evidence of a hypernova imprint in any of the Milky Way satellite galaxies. In the Milky Way halo,
there are two other possible detections reported in the literature, 
albeit each displaying rather different abundance signatures.
We discuss these in turn below.

Recently it was argued that the CEMP-no star HE~1327$-$2326 in the
Milky Way halo shows an imprint of a SN with 
rather high explosion energy, $E=5\times10^{51}$\,erg \citep{Ezzeddine19}.
However, their models depend on asymmetrical effects, which are still
poorly understood; these effects moreover introduce three additional free
parameters to be fitted.  Furthermore, more recent models struggle to
produce the large [C/Fe] observed in HE 1327$-$2326 \citep{Grimmett21}. 
In contrast, the abundance pattern of AS0039 can be matched by a simple
spherical hypernova, without having to invoke any additional physics and
free parameters.

Even more recently, \citet{Placco21} presented results for 
SPLUS J210428.01-004934.2. They demonstrated that 
its abundance pattern is best fit with a progenitor of $30$\,M$_\odot$,
that is $\sim$50\% larger than that of AS0039, but with the same 
explosion energy ($E=10\times10^{51}$\,erg).
Interestingly, although J210428.01-004934.2 is also somewhat low in C, 
$\rm {[C/Fe]_{LTE}=-0.1}$,
it displays $\rm [Mg/Ca,Ti]_{LTE}$ ratios that are close to zero, in stark contrast 
to the atypically low values measured in AS0039 (Fig.~\ref{fig:MgCaTi}). 
Clearly the abundance signatures of PopIII hypernovae have a 
strong mass dependence, 
and further observations are needed to fully understand the diversity of
the first generations of stars.

\section{Conclusion} \label{sec:conclusion}

New ESO VLT/X-Shooter observations have revealed that at $\rm [Fe/H]_{\rm LTE}=-4.11$ ($\rm [Fe/H]_{NLTE}=-3.95$), with unusually low $\rm [Mg/Fe]_{NLTE}=-0.16$, and $\rm [C/Fe]_{corr}=-0.29$, the star AS0039
in the Sculptor dwarf spheroidal is the most metal-poor star that has been discovered in an external galaxy. Furthermore, it has the currently lowest detected C abundance, $A(C)=+3.60$, in any galaxy.
In addition to the overall low metallicity, this star has 
a unique abundance pattern, with particularly low ratios of hydrostatic-to-explosive $\alpha$-elements ($\rm [Mg/Ca,Ti]<-0.6$). 
The detailed non-LTE abundance pattern of AS0039 can best be explained with a Population III progenitor of $M=21$\,M$_\odot$, and 
high explosion energy $E=10\times10^{51}$~erg, making AS0039 one of the first observational evidence of a zero-metallicity hypernova. 

The number of known stars in the Milky Way, at $\rm [Fe/H]<-3.5$, is almost an order of magnitude higher than in the dwarf galaxy satellites \citep{Suda08}. Given the available data, it is therefore unlikely merely a coincidence that one of two known first descendant of a
zero-metallicity hypernova is found in Sculptor. First of all, the galaxy formed as much as 80\% of its
stars 12-14 Gyr ago \citep{Bettinelli19}, and thus it gives an
unobscured view to the earliest star formation. Secondly, Sculptor may lie in a sweet spot in terms of mass. Compared to UFD galaxies it
is relatively massive \citep{Battaglia08}, with $\rm M_{tot}\gtrsim 4 \times 10^8$. If a rare event such
as hypernovae were to occur in a much smaller UFD galaxy, it
would likely be too small to retain the yields of such an energetic event \citep{Cooke14}. 
At the same time, given the only modest size and star formation rate of Sculptor, the imprints of hypernovae will not be lost or 
rapidly diluted by subsequent star formation events.
The Sculptor dwarf
spheroidal galaxy may thus be the ideal system for further discoveries.

Until now, the observational evidence for Population~III
stars has been completely
dominated by the imprints of faint supernovae seen in CEMP-no
stars. These low-energy supernovae, however, only occupy a limited portion of
the parameter space that has been theoretically predicted for Population~III stars.
With the discovery of AS0039, we are for the first time expanding this view,
opening a window into the investigation of the most energetic primordial
supernovae, i.e. Population~III hypernovae. Combined with advances in theoretical
simulations of early galaxy formation and improved data from upcoming large
spectroscopic surveys (e.g.,~\citealt{Christlieb19}), this discovery will enable us to assemble the full
picture of the elusive nature of the first stars in the Universe.

\acknowledgments{
These results are based on VLT/FLAMES and VLT/X-Shooter observations collected at the European Organisation for Astronomical Research (ESO) in the Southern Hemisphere under programme ESO ID 0102.B-0786.
This project has received funding from the European Research Council (ERC) under the European Union's Horizon 2020 research and innovation programme (grant agreement No.~804240).
A.M.A. gratefully acknowledges support from the Swedish Research Council (VR~2020-03940).
G.B. acknowledges financial support through the grant (AEI/FEDER, UE) AYA2017-89076-P, as well as by the Ministerio de Ciencia, Innovación y Universidades (MCIU), through the State Budget and by the Consejería de Economía, Industria, Comercio y Conocimiento of the Canary Islands Autonomous Community, through the Regional Budget. E.S. acknowledges funding through VIDI grant ``Pushing Galactic Archaeology to its limits" (with project number VI.Vidi.193.093) which is funded by the Dutch Research Council (NWO).
This work was supported by computational resources provided by 
the Australian Government through the 
National Computational Infrastructure (NCI)
under the National Computational Merit Allocation Scheme (NCMAS)
and Australian National University Merit Allocation Scheme (ANUMAS).}

\vspace{5mm}
\facilities{ESO VLT/FLAMES, ESO VLT/X-SHOOTER, {\it Gaia}.}

\bibliography{heimildir,heimildir_data}

\begin{thebibliography}{}
\expandafter\ifx\csname natexlab\endcsname\relax\def\natexlab#1{#1}\fi
\providecommand{\url}[1]{\href{#1}{#1}}
\providecommand{\dodoi}[1]{doi:~\href{http://doi.org/#1}{\nolinkurl{#1}}}
\providecommand{\doeprint}[1]{\href{http://ascl.net/#1}{\nolinkurl{http://ascl.net/#1}}}
\providecommand{\doarXiv}[1]{\href{https://arxiv.org/abs/#1}{\nolinkurl{https://arxiv.org/abs/#1}}}

\bibitem[{{Amarsi} \& {Asplund}(2017)}]{Amarsi17}
{Amarsi}, A.~M., \& {Asplund}, M. 2017, MNRAS, 464, 264,
  \dodoi{10.1093/mnras/stw2445}

\bibitem[{{Amarsi} {et~al.}(2016){Amarsi}, {Lind}, {Asplund}, {Barklem}, \&
  {Collet}}]{AmarsiFe}
{Amarsi}, A.~M., {Lind}, K., {Asplund}, M., {Barklem}, P.~S., \& {Collet}, R.
  2016, MNRAS, 463, 1518, \dodoi{10.1093/mnras/stw2077}

\bibitem[{{Amarsi} {et~al.}(2019){Amarsi}, {Nissen}, \&
  {Sk{\'u}lad{\'o}ttir}}]{Amarsi19}
{Amarsi}, A.~M., {Nissen}, P.~E., \& {Sk{\'u}lad{\'o}ttir}, {\'A}. 2019,
  Astron. \& Astrophys., 630, A104, \dodoi{10.1051/0004-6361/201936265}

\bibitem[{{Amarsi} {et~al.}(2018){Amarsi}, {Nordlander}, {Barklem}, {Asplund},
  {Collet}, \& {Lind}}]{Amarsi18}
{Amarsi}, A.~M., {Nordlander}, T., {Barklem}, P.~S., {et~al.} 2018, Astron. \&
  Astrophys., 615, A139, \dodoi{10.1051/0004-6361/201732546}

\bibitem[{{Amarsi} {et~al.}(2020){Amarsi}, {Lind}, {Osorio}, {Nordlander},
  {Bergemann}, {Reggiani}, {Wang}, {Buder}, {Asplund}, {Barklem}, {Wehrhahn},
  {Sk{\'u}lad{\'o}ttir}, {Kobayashi}, {Karakas}, {Gao}, {Bland-Hawthorn}, {de
  Silva}, {Kos}, {Lewis}, {Martell}, {Sharma}, {Simpson}, {Zucker},
  {{\v{C}}otar}, {Horner}, \& {Galah Collaboration}}]{Amarsi20}
{Amarsi}, A.~M., {Lind}, K., {Osorio}, Y., {et~al.} 2020, Astron. \&
  Astrophys., 642, A62, \dodoi{10.1051/0004-6361/202038650}

\bibitem[{{Andrievsky} {et~al.}(2010){Andrievsky}, {Spite}, {Korotin}, {Spite},
  {Bonifacio}, {Cayrel}, {Fran{\c{c}}ois}, \& {Hill}}]{Andrievsky10}
{Andrievsky}, S.~M., {Spite}, M., {Korotin}, S.~A., {et~al.} 2010, Astron. \&
  Astrophys., 509, A88, \dodoi{10.1051/0004-6361/200913223}

\bibitem[{{Asplund} {et~al.}(2021){Asplund}, {Amarsi}, \&
  {Grevesse}}]{Asplund21}
{Asplund}, M., {Amarsi}, A.~M., \& {Grevesse}, N. 2021, arXiv e-prints,
  arXiv:2105.01661.
\newblock \doarXiv{2105.01661}

\bibitem[{{Barklem} {et~al.}(2021){Barklem}, {Amarsi}, {Grumer}, {Eklund},
  {Ros{\'e}n}, {Ji}, {Cederquist}, {Zettergren}, \& {Schmidt}}]{Barklem21}
{Barklem}, P.~S., {Amarsi}, A.~M., {Grumer}, J., {et~al.} 2021, Astrophys. J.,
  908, 245, \dodoi{10.3847/1538-4357/abd5bd}

\bibitem[{{Battaglia} {et~al.}(2008){Battaglia}, {Helmi}, {Tolstoy}, {Irwin},
  {Hill}, \& {Jablonka}}]{Battaglia08}
{Battaglia}, G., {Helmi}, A., {Tolstoy}, E., {et~al.} 2008, Astrophys. J. L.,
  681, L13, \dodoi{10.1086/590179}

\bibitem[{{Bergemann}(2011)}]{Bergemann11}
{Bergemann}, M. 2011, MNRAS, 413, 2184,
  \dodoi{10.1111/j.1365-2966.2011.18295.x}

\bibitem[{{Bergemann} \& {Cescutti}(2010)}]{Bergemann10}
{Bergemann}, M., \& {Cescutti}, G. 2010, Astron. \& Astrophys., 522, A9,
  \dodoi{10.1051/0004-6361/201014250}

\bibitem[{{Bergemann} {et~al.}(2012){Bergemann}, {Hansen}, {Bautista}, \&
  {Ruchti}}]{BergemannSr}
{Bergemann}, M., {Hansen}, C.~J., {Bautista}, M., \& {Ruchti}, G. 2012, Astron.
  \& Astrophys., 546, A90, \dodoi{10.1051/0004-6361/201219406}

\bibitem[{{Bergemann} {et~al.}(2010){Bergemann}, {Pickering}, \&
  {Gehren}}]{BergemannCo}
{Bergemann}, M., {Pickering}, J.~C., \& {Gehren}, T. 2010, MNRAS, 401, 1334,
  \dodoi{10.1111/j.1365-2966.2009.15736.x}

\bibitem[{{Bergemann} {et~al.}(2019){Bergemann}, {Gallagher}, {Eitner},
  {Bautista}, {Collet}, {Yakovleva}, {Mayriedl}, {Plez}, {Carlsson},
  {Leenaarts}, {Belyaev}, \& {Hansen}}]{BergemannMn}
{Bergemann}, M., {Gallagher}, A.~J., {Eitner}, P., {et~al.} 2019, Astron. \&
  Astrophys., 631, A80, \dodoi{10.1051/0004-6361/201935811}

\bibitem[{{Bettinelli} {et~al.}(2019){Bettinelli}, {Hidalgo}, {Cassisi},
  {Aparicio}, {Piotto}, {Valdes}, \& {Walker}}]{Bettinelli19}
{Bettinelli}, M., {Hidalgo}, S.~L., {Cassisi}, S., {et~al.} 2019, MNRAS, 487,
  5862, \dodoi{10.1093/mnras/stz1679}

\bibitem[{{Caffau} {et~al.}(2011){Caffau}, {Bonifacio}, {Fran{\c c}ois},
  {Sbordone}, {Monaco}, {Spite}, {Spite}, {Ludwig}, {Cayrel}, {Zaggia},
  {Hammer}, {Randich}, {Molaro}, \& {Hill}}]{Caffau11}
{Caffau}, E., {Bonifacio}, P., {Fran{\c c}ois}, P., {et~al.} 2011, Nature, 477,
  67, \dodoi{10.1038/nature10377}

\bibitem[{{Cayrel} {et~al.}(2004){Cayrel}, {Depagne}, {Spite}, {Hill}, {Spite},
  {Fran{\c{c}}ois}, {Plez}, {Beers}, {Primas}, {Andersen}, {Barbuy},
  {Bonifacio}, {Molaro}, \& {Nordstr{\"o}m}}]{Cayrel04}
{Cayrel}, R., {Depagne}, E., {Spite}, M., {et~al.} 2004, Astron. \& Astrophys.,
  416, 1117, \dodoi{10.1051/0004-6361:20034074}

\bibitem[{{Chiti} {et~al.}(2018){Chiti}, {Frebel}, {Ji}, {Jerjen}, {Kim}, \&
  {Norris}}]{Chiti18tucII}
{Chiti}, A., {Frebel}, A., {Ji}, A.~P., {et~al.} 2018, Astrophys. J., 857, 74,
  \dodoi{10.3847/1538-4357/aab4fc}

\bibitem[{{Christlieb} {et~al.}(2019){Christlieb}, {Battistini}, {Bonifacio},
  {Caffau}, {Ludwig}, {Asplund}, {Barklem}, {Bergemann}, {Church}, {Feltzing},
  {Ford}, {Grebel}, {Hansen}, {Helmi}, {Kordopatis}, {Kovalev}, {Korn}, {Lind},
  {Quirrenbach}, {Rybizki}, {Sk{\'u}lad{\'o}ttir}, \&
  {Starkenburg}}]{Christlieb19}
{Christlieb}, N., {Battistini}, C., {Bonifacio}, P., {et~al.} 2019, The
  Messenger, 175, 26, \dodoi{10.18727/0722-6691/5121}

\bibitem[{{Cohen} \& {Huang}(2009)}]{Cohen09}
{Cohen}, J.~G., \& {Huang}, W. 2009, Astrophys. J., 701, 1053,
  \dodoi{10.1088/0004-637X/701/2/1053}

\bibitem[{{Cohen} \& {Huang}(2010)}]{Cohen10}
---. 2010, Astrophys. J., 719, 931, \dodoi{10.1088/0004-637X/719/1/931}

\bibitem[{{Cooke} \& {Madau}(2014)}]{Cooke14}
{Cooke}, R.~J., \& {Madau}, P. 2014, Astrophys. J., 791, 116,
  \dodoi{10.1088/0004-637X/791/2/116}

\bibitem[{{Ezzeddine} {et~al.}(2019){Ezzeddine}, {Frebel}, {Roederer},
  {Tominaga}, {Tumlinson}, {Ishigaki}, {Nomoto}, {Placco}, \&
  {Aoki}}]{Ezzeddine19}
{Ezzeddine}, R., {Frebel}, A., {Roederer}, I.~U., {et~al.} 2019, \apj, 876, 97,
  \dodoi{10.3847/1538-4357/ab14e7}

\bibitem[{{Fran{\c{c}}ois} {et~al.}(2016){Fran{\c{c}}ois}, {Monaco},
  {Bonifacio}, {Moni Bidin}, {Geisler}, \& {Sbordone}}]{Francois16}
{Fran{\c{c}}ois}, P., {Monaco}, L., {Bonifacio}, P., {et~al.} 2016, Astron. \&
  Astrophys., 588, A7, \dodoi{10.1051/0004-6361/201527181}

\bibitem[{{Frebel} {et~al.}(2010{\natexlab{a}}){Frebel}, {Kirby}, \&
  {Simon}}]{Frebel10}
{Frebel}, A., {Kirby}, E.~N., \& {Simon}, J.~D. 2010{\natexlab{a}}, Nature,
  464, 72, \dodoi{10.1038/nature08772}

\bibitem[{{Frebel} {et~al.}(2016){Frebel}, {Norris}, {Gilmore}, \&
  {Wyse}}]{Frebel16}
{Frebel}, A., {Norris}, J.~E., {Gilmore}, G., \& {Wyse}, R. F.~G. 2016,
  Astrophys. J., 826, 110, \dodoi{10.3847/0004-637X/826/2/110}

\bibitem[{{Frebel} {et~al.}(2010{\natexlab{b}}){Frebel}, {Simon}, {Geha}, \&
  {Willman}}]{Frebel10b}
{Frebel}, A., {Simon}, J.~D., {Geha}, M., \& {Willman}, B. 2010{\natexlab{b}},
  Astrophys. J., 708, 560, \dodoi{10.1088/0004-637X/708/1/560}

\bibitem[{{Frebel} {et~al.}(2014){Frebel}, {Simon}, \& {Kirby}}]{Frebel14}
{Frebel}, A., {Simon}, J.~D., \& {Kirby}, E.~N. 2014, Astrophys. J., 786, 74,
  \dodoi{10.1088/0004-637X/786/1/74}

\bibitem[{{Gaia Collaboration} {et~al.}(2020){Gaia Collaboration}, {Brown},
  {Vallenari}, {Prusti}, {de Bruijne}, {Babusiaux}, \& {Biermann}}]{Gaia20}
{Gaia Collaboration}, {Brown}, A.~G.~A., {Vallenari}, A., {et~al.} 2020, arXiv
  e-prints, arXiv:2012.01533.
\newblock \doarXiv{2012.01533}

\bibitem[{{Gilmore} {et~al.}(2013){Gilmore}, {Norris}, {Monaco}, {Yong},
  {Wyse}, \& {Geisler}}]{Gilmore13}
{Gilmore}, G., {Norris}, J.~E., {Monaco}, L., {et~al.} 2013, Astrophys. J.,
  763, 61, \dodoi{10.1088/0004-637X/763/1/61}

\bibitem[{{Grimmett} {et~al.}(2021){Grimmett}, {M{\"u}ller}, {Heger},
  {Banerjee}, \& {Obergaulinger}}]{Grimmett21}
{Grimmett}, J.~J., {M{\"u}ller}, B., {Heger}, A., {Banerjee}, P., \&
  {Obergaulinger}, M. 2021, \mnras, 501, 2764, \dodoi{10.1093/mnras/staa3819}

\bibitem[{{Gustafsson} {et~al.}(2008){Gustafsson}, {Edvardsson}, {Eriksson},
  {J{\o}rgensen}, {Nordlund}, \& {Plez}}]{Gustafsson08}
{Gustafsson}, B., {Edvardsson}, B., {Eriksson}, K., {et~al.} 2008, Astron. \&
  Astrophys., 486, 951, \dodoi{10.1051/0004-6361:200809724}

\bibitem[{{Heger} \& {Woosley}(2010)}]{HegerWoosley10}
{Heger}, A., \& {Woosley}, S.~E. 2010, Astrophys. J., 724, 341,
  \dodoi{10.1088/0004-637X/724/1/341}

\bibitem[{{Hill} {et~al.}(2019){Hill}, {Sk{\'u}lad{\'o}ttir}, {Tolstoy},
  {Venn}, {Shetrone}, {Jablonka}, {Primas}, {Battaglia}, {de Boer},
  {Fran{\c{c}}ois}, {Helmi}, {Kaufer}, {Letarte}, {Starkenburg}, \&
  {Spite}}]{Hill19}
{Hill}, V., {Sk{\'u}lad{\'o}ttir}, {\'A}., {Tolstoy}, E., {et~al.} 2019,
  Astron. \& Astrophys., 626, A15, \dodoi{10.1051/0004-6361/201833950}

\bibitem[{{Iwamoto} {et~al.}(2005){Iwamoto}, {Umeda}, {Tominaga}, {Nomoto}, \&
  {Maeda}}]{Iwamoto05}
{Iwamoto}, N., {Umeda}, H., {Tominaga}, N., {Nomoto}, K., \& {Maeda}, K. 2005,
  Science, 309, 451, \dodoi{10.1126/science.1112997}

\bibitem[{{Jablonka} {et~al.}(2015){Jablonka}, {North}, {Mashonkina}, {Hill},
  {Revaz}, {Shetrone}, {Starkenburg}, {Irwin}, {Tolstoy}, {Battaglia}, {Venn},
  {Helmi}, {Primas}, \& {Fran{\c{c}}ois}}]{Jablonka15}
{Jablonka}, P., {North}, P., {Mashonkina}, L., {et~al.} 2015, Astron. \&
  Astrophys., 583, A67, \dodoi{10.1051/0004-6361/201525661}

\bibitem[{{Ji} {et~al.}(2016{\natexlab{a}}){Ji}, {Frebel}, {Ezzeddine}, \&
  {Casey}}]{Ji16tucII}
{Ji}, A.~P., {Frebel}, A., {Ezzeddine}, R., \& {Casey}, A.~R.
  2016{\natexlab{a}}, Astrophys. J. L., 832, L3,
  \dodoi{10.3847/2041-8205/832/1/L3}

\bibitem[{{Ji} {et~al.}(2016{\natexlab{b}}){Ji}, {Frebel}, {Simon}, \&
  {Chiti}}]{Ji16retII}
{Ji}, A.~P., {Frebel}, A., {Simon}, J.~D., \& {Chiti}, A. 2016{\natexlab{b}},
  Astrophys. J., 830, 93, \dodoi{10.3847/0004-637X/830/2/93}

\bibitem[{{Ji} {et~al.}(2016{\natexlab{c}}){Ji}, {Frebel}, {Simon}, \&
  {Geha}}]{Ji16booII}
{Ji}, A.~P., {Frebel}, A., {Simon}, J.~D., \& {Geha}, M. 2016{\natexlab{c}},
  Astrophys. J., 817, 41, \dodoi{10.3847/0004-637X/817/1/41}

\bibitem[{{Ji} {et~al.}(2019){Ji}, {Simon}, {Frebel}, {Venn}, \&
  {Hansen}}]{Ji19}
{Ji}, A.~P., {Simon}, J.~D., {Frebel}, A., {Venn}, K.~A., \& {Hansen}, T.~T.
  2019, Astrophys. J., 870, 83, \dodoi{10.3847/1538-4357/aaf3bb}

\bibitem[{{Kirby} \& {Cohen}(2012)}]{KirbyCohen12}
{Kirby}, E.~N., \& {Cohen}, J.~G. 2012, Astronomical J., 144, 168,
  \dodoi{10.1088/0004-6256/144/6/168}

\bibitem[{{Kirby} {et~al.}(2009){Kirby}, {Guhathakurta}, {Bolte}, {Sneden}, \&
  {Geha}}]{Kirby09}
{Kirby}, E.~N., {Guhathakurta}, P., {Bolte}, M., {Sneden}, C., \& {Geha}, M.~C.
  2009, Astrophys. J., 705, 328, \dodoi{10.1088/0004-637X/705/1/328}

\bibitem[{{Kobayashi} {et~al.}(2020){Kobayashi}, {Karakas}, \&
  {Lugaro}}]{Kobayashi20}
{Kobayashi}, C., {Karakas}, A.~I., \& {Lugaro}, M. 2020, Astrophys. J., 900,
  179, \dodoi{10.3847/1538-4357/abae65}

\bibitem[{Kovalev(2019)}]{Kovalev19}
Kovalev, M. 2019, PhD thesis, Universit{\"a}t Heidelberg

\bibitem[{{Lai} {et~al.}(2008){Lai}, {Bolte}, {Johnson}, {Lucatello}, {Heger},
  \& {Woosley}}]{Lai08}
{Lai}, D.~K., {Bolte}, M., {Johnson}, J.~A., {et~al.} 2008, Astrophys. J., 681,
  1524, \dodoi{10.1086/588811}

\bibitem[{{Lee} {et~al.}(2013){Lee}, {Beers}, {Masseron}, {Plez}, {Rockosi},
  {Sobeck}, {Yanny}, {Lucatello}, {Sivarani}, {Placco}, \& {Carollo}}]{Lee13}
{Lee}, Y.~S., {Beers}, T.~C., {Masseron}, T., {et~al.} 2013, Astron. J., 146,
  132, \dodoi{10.1088/0004-6256/146/5/132}

\bibitem[{{Leenaarts} \& {Carlsson}(2009)}]{Leenaarts09}
{Leenaarts}, J., \& {Carlsson}, M. 2009, in Astronomical Society of the Pacific
  Conference Series, Vol. 415, The Second Hinode Science Meeting: Beyond
  Discovery-Toward Understanding, ed. B.~{Lites}, M.~{Cheung}, T.~{Magara},
  J.~{Mariska}, \& K.~{Reeves}, 87

\bibitem[{{Lind} {et~al.}(2011){Lind}, {Asplund}, {Barklem}, \&
  {Belyaev}}]{Lind11}
{Lind}, K., {Asplund}, M., {Barklem}, P.~S., \& {Belyaev}, A.~K. 2011, Astron.
  \& Astrophys., 528, A103, \dodoi{10.1051/0004-6361/201016095}

\bibitem[{{Mucciarelli} \& {Bellazzini}(2020)}]{Mucciarelli20}
{Mucciarelli}, A., \& {Bellazzini}, M. 2020, Research Notes of the American
  Astronomical Society, 4, 52, \dodoi{10.3847/2515-5172/ab8820}

\bibitem[{{Nordlander} \& {Lind}(2017)}]{Nordlander17}
{Nordlander}, T., \& {Lind}, K. 2017, Astron. \& Astrophys., 607, A75,
  \dodoi{10.1051/0004-6361/201730427}

\bibitem[{{Norris} {et~al.}(2010{\natexlab{a}}){Norris}, {Gilmore}, {Wyse},
  {Yong}, \& {Frebel}}]{Norris10seg}
{Norris}, J.~E., {Gilmore}, G., {Wyse}, R. F.~G., {Yong}, D., \& {Frebel}, A.
  2010{\natexlab{a}}, Astrophys. J. L., 722, L104,
  \dodoi{10.1088/2041-8205/722/1/L104}

\bibitem[{{Norris} \& {Yong}(2019)}]{Norris19}
{Norris}, J.~E., \& {Yong}, D. 2019, Astrophys. J., 879, 37,
  \dodoi{10.3847/1538-4357/ab1f84}

\bibitem[{{Norris} {et~al.}(2010{\natexlab{b}}){Norris}, {Yong}, {Gilmore}, \&
  {Wyse}}]{Norris10boo}
{Norris}, J.~E., {Yong}, D., {Gilmore}, G., \& {Wyse}, R. F.~G.
  2010{\natexlab{b}}, Astrophys. J., 711, 350,
  \dodoi{10.1088/0004-637X/711/1/350}

\bibitem[{{Norris} {et~al.}(2013){Norris}, {Yong}, {Bessell}, {Christlieb},
  {Asplund}, {Gilmore}, {Wyse}, {Beers}, {Barklem}, {Frebel}, \&
  {Ryan}}]{Norris13}
{Norris}, J.~E., {Yong}, D., {Bessell}, M.~S., {et~al.} 2013, Astrophys. J.,
  762, 28, \dodoi{10.1088/0004-637X/762/1/28}

\bibitem[{{Placco} {et~al.}(2014){Placco}, {Frebel}, {Beers}, \&
  {Stancliffe}}]{Placco14}
{Placco}, V.~M., {Frebel}, A., {Beers}, T.~C., \& {Stancliffe}, R.~J. 2014,
  Astrophys. J., 797, 21, \dodoi{10.1088/0004-637X/797/1/21}

\bibitem[{{Placco} {et~al.}(2021){Placco}, {Roederer}, {Lee},
  {Almeida-Fernandes}, {Herpich}, {Perottoni}, {Schoenell}, {Ribeiro}, \&
  {Kanaan}}]{Placco21}
{Placco}, V.~M., {Roederer}, I.~U., {Lee}, Y.~S., {et~al.} 2021, \apjl, 912,
  L32, \dodoi{10.3847/2041-8213/abf93d}

\bibitem[{{Plez}(2012)}]{Plez12}
{Plez}, B. 2012, {Turbospectrum: Code for spectral synthesis}.
\newblock \doeprint{1205.004}

\bibitem[{{Roederer} \& {Kirby}(2014)}]{RoedererKirby14}
{Roederer}, I.~U., \& {Kirby}, E.~N. 2014, MNRAS, 440, 2665,
  \dodoi{10.1093/mnras/stu491}

\bibitem[{{Roederer} {et~al.}(2014){Roederer}, {Preston}, {Thompson},
  {Shectman}, {Sneden}, {Burley}, \& {Kelson}}]{Roederer14}
{Roederer}, I.~U., {Preston}, G.~W., {Thompson}, I.~B., {et~al.} 2014,
  Astronomical J., 147, 136, \dodoi{10.1088/0004-6256/147/6/136}

\bibitem[{{Rossi} {et~al.}(2021){Rossi}, {Salvadori}, \&
  {Sk{\'u}lad{\'o}ttir}}]{Rossi21}
{Rossi}, M., {Salvadori}, S., \& {Sk{\'u}lad{\'o}ttir}, {\'A}. 2021, MNRAS,
  \dodoi{10.1093/mnras/stab821}

\bibitem[{{Salvadori} {et~al.}(2019){Salvadori}, {Bonifacio}, {Caffau},
  {Korotin}, {Andreevsky}, {Spite}, \& {Sk{\'u}lad{\'o}ttir}}]{Salvadori19}
{Salvadori}, S., {Bonifacio}, P., {Caffau}, E., {et~al.} 2019, MNRAS, 487,
  4261, \dodoi{10.1093/mnras/stz1464}

\bibitem[{{Shetrone} {et~al.}(2001){Shetrone}, {C{\^o}t{\'e}}, \&
  {Sargent}}]{Shetrone01}
{Shetrone}, M.~D., {C{\^o}t{\'e}}, P., \& {Sargent}, W.~L.~W. 2001, Astrophys.
  J., 548, 592, \dodoi{10.1086/319022}

\bibitem[{{Simon} {et~al.}(2010){Simon}, {Frebel}, {McWilliam}, {Kirby}, \&
  {Thompson}}]{Simon10}
{Simon}, J.~D., {Frebel}, A., {McWilliam}, A., {Kirby}, E.~N., \& {Thompson},
  I.~B. 2010, Astrophys. J., 716, 446, \dodoi{10.1088/0004-637X/716/1/446}

\bibitem[{{Simon} {et~al.}(2015){Simon}, {Jacobson}, {Frebel}, {Thompson},
  {Adams}, \& {Shectman}}]{Simon15}
{Simon}, J.~D., {Jacobson}, H.~R., {Frebel}, A., {et~al.} 2015, Astrophys. J.,
  802, 93, \dodoi{10.1088/0004-637X/802/2/93}

\bibitem[{{Sk{\'u}lad{\'o}ttir}
  {et~al.}(2015{\natexlab{a}}){Sk{\'u}lad{\'o}ttir}, {Andrievsky}, {Tolstoy},
  {Hill}, {Salvadori}, {Korotin}, \& {Pettini}}]{Skuladottir15b}
{Sk{\'u}lad{\'o}ttir}, {\'A}., {Andrievsky}, S.~M., {Tolstoy}, E., {et~al.}
  2015{\natexlab{a}}, Astron. \& Astrophys., 580, A129,
  \dodoi{10.1051/0004-6361/201525956}

\bibitem[{{Sk{\'u}lad{\'o}ttir} {et~al.}(2017){Sk{\'u}lad{\'o}ttir}, {Tolstoy},
  {Salvadori}, {Hill}, \& {Pettini}}]{Skuladottir17}
{Sk{\'u}lad{\'o}ttir}, {\'A}., {Tolstoy}, E., {Salvadori}, S., {Hill}, V., \&
  {Pettini}, M. 2017, Astron. \& Astrophys., 606, A71,
  \dodoi{10.1051/0004-6361/201731158}

\bibitem[{{Sk{\'u}lad{\'o}ttir}
  {et~al.}(2015{\natexlab{b}}){Sk{\'u}lad{\'o}ttir}, {Tolstoy}, {Salvadori},
  {Hill}, {Pettini}, {Shetrone}, \& {Starkenburg}}]{Skuladottir15}
{Sk{\'u}lad{\'o}ttir}, {\'A}., {Tolstoy}, E., {Salvadori}, S., {et~al.}
  2015{\natexlab{b}}, Astron. \& Astrophys., 574, A129,
  \dodoi{10.1051/0004-6361/201424782}

\bibitem[{{Starkenburg} {et~al.}(2013){Starkenburg}, {Hill}, {Tolstoy},
  {Fran{\c{c}}ois}, {Irwin}, {Boschman}, {Venn}, {de Boer}, {Lemasle},
  {Jablonka}, {Battaglia}, {Groot}, \& {Kaper}}]{Starkenburg13}
{Starkenburg}, E., {Hill}, V., {Tolstoy}, E., {et~al.} 2013, Astron. \&
  Astrophys., 549, A88, \dodoi{10.1051/0004-6361/201220349}

\bibitem[{{Starkenburg} {et~al.}(2018){Starkenburg}, {Aguado}, {Bonifacio},
  {Caffau}, {Jablonka}, {Lardo}, {Martin}, {S{\'a}nchez-Janssen}, {Sestito},
  {Venn}, {Youakim}, {Allende Prieto}, {Arentsen}, {Gentile}, {Gonz{\'a}lez
  Hern{\'a}ndez}, {Kielty}, {Koppelman}, {Longeard}, {Tolstoy}, {Carlberg},
  {C{\^o}t{\'e}}, {Fouesneau}, {Hill}, {McConnachie}, \&
  {Navarro}}]{Starkenburg18}
{Starkenburg}, E., {Aguado}, D.~S., {Bonifacio}, P., {et~al.} 2018, \mnras,
  481, 3838, \dodoi{10.1093/mnras/sty2276}

\bibitem[{{Suda} {et~al.}(2008){Suda}, {Katsuta}, {Yamada}, {Suwa}, {Ishizuka},
  {Komiya}, {Sorai}, {Aikawa}, \& {Fujimoto}}]{Suda08}
{Suda}, T., {Katsuta}, Y., {Yamada}, S., {et~al.} 2008, PASJ, 60, 1159,
  \dodoi{10.1093/pasj/60.5.1159}

\bibitem[{{Tafelmeyer} {et~al.}(2010){Tafelmeyer}, {Jablonka}, {Hill},
  {Shetrone}, {Tolstoy}, {Irwin}, {Battaglia}, {Helmi}, {Starkenburg}, {Venn},
  {Abel}, {Francois}, {Kaufer}, {North}, {Primas}, \&
  {Szeifert}}]{Tafelmeyer10}
{Tafelmeyer}, M., {Jablonka}, P., {Hill}, V., {et~al.} 2010, Astron. \&
  Astrophys., 524, A58, \dodoi{10.1051/0004-6361/201014733}

\bibitem[{{Theler} {et~al.}(2020){Theler}, {Jablonka}, {Lucchesi}, {Lardo},
  {North}, {Irwin}, {Battaglia}, {Hill}, {Tolstoy}, {Venn}, {Helmi}, {Kaufer},
  {Primas}, \& {Shetrone}}]{Theler20}
{Theler}, R., {Jablonka}, P., {Lucchesi}, R., {et~al.} 2020, Astron. \&
  Astrophys., 642, A176, \dodoi{10.1051/0004-6361/201937146}

\bibitem[{{Venn} {et~al.}(2017){Venn}, {Starkenburg}, {Malo}, {Martin}, \&
  {Laevens}}]{Venn17}
{Venn}, K.~A., {Starkenburg}, E., {Malo}, L., {Martin}, N., \& {Laevens},
  B.~P.~M. 2017, MNRAS, 466, 3741, \dodoi{10.1093/mnras/stw3198}

\bibitem[{{Wang} {et~al.}(2021){Wang}, {Nordlander}, {Asplund}, {Amarsi},
  {Lind}, \& {Zhou}}]{Wang21}
{Wang}, E.~X., {Nordlander}, T., {Asplund}, M., {et~al.} 2021, MNRAS, 500,
  2159, \dodoi{10.1093/mnras/staa3381}

\bibitem[{{Yong} {et~al.}(2013){Yong}, {Norris}, {Bessell}, {Christlieb},
  {Asplund}, {Beers}, {Barklem}, {Frebel}, \& {Ryan}}]{Yong13}
{Yong}, D., {Norris}, J.~E., {Bessell}, M.~S., {et~al.} 2013, Astrophys. J.,
  762, 26, \dodoi{10.1088/0004-637X/762/1/26}

\bibitem[{{Yoon} {et~al.}(2018){Yoon}, {Beers}, {Dietz}, {Lee}, {Placco}, {Da
  Costa}, {Keller}, {Owen}, \& {Sharma}}]{Yoon18}
{Yoon}, J., {Beers}, T.~C., {Dietz}, S., {et~al.} 2018, \apj, 861, 146,
  \dodoi{10.3847/1538-4357/aaccea}

\bibitem[{{Zhang} {et~al.}(2014){Zhang}, {Gehren}, \& {Zhao}}]{Zhang14}
{Zhang}, H.~W., {Gehren}, T., \& {Zhao}, G. 2014, in Setting the scene for Gaia
  and LAMOST, ed. S.~{Feltzing}, G.~{Zhao}, N.~A. {Walton}, \& P.~{Whitelock},
  Vol. 298, 453--453, \dodoi{10.1017/S1743921313007187}

\end{thebibliography}

\clearpage
\pagebreak

\appendix


\section{Literature compilation} \label{app:literature}
The measured abundance ratios of [Mg/Ca] and [Mg/Ti] in AS0039 are compared to literature values in Fig.~2. Typically only LTE abundance analysis is provided in the literature \citep{Suda08}. However, NLTE effects for similar stars (in terms of $T_{\rm eff}$, $\log g$, and [Fe/H]) are expected to be in the same direction and have comparable magnitudes. All existing stellar abundance measurements in dwarf galaxies are for RGB stars, as main-sequence stars are too faint for such analysis to be feasible. In Fig.~2 we therefore only include RGB stars in the Milky Way to ensure a fair comparison. Where provided, literature abundances based on Fe~I and Ti~II are chosen, to be as consistent to the analysis of AS0039 as possible. \\ \indent
To show the uniqueness of [Mg/Ca] and [Mg/Ti] in AS0039, our literature compilation was quite extensive, especially in relation to other dwarf galaxies. Fig.~2 includes classical 1D LTE abundances for RGB stars in Sculptor \citep{Starkenburg13,Jablonka15,Tafelmeyer10,Frebel10,Simon15}, other dSph galaxies \citep{Tafelmeyer10,Theler20,Shetrone01,Cohen09,Cohen10,KirbyCohen12}, the UFD galaxies \citep{Chiti18tucII,Francois16,Frebel10b,Frebel14,Frebel16,Norris10boo,Gilmore13,Ji16booII,Ji16retII,Ji16tucII,Ji19,Norris10seg,
		RoedererKirby14,Simon10,Venn17}, and C-normal RGB stars in the Milky Way \citep{Cayrel04,Lai08,Yong13,Roederer14}. All literature data shown or discussed in this paper have been put to the same solar abundance scale \citep{Asplund21}.

The general abundance pattern of AS0039, compared to other Sculptor stars, and the Milky Way, is shown in Fig.~4. 

        \begin{figure}[!h] \label{fig:abuall}
   \centering
   \includegraphics[width=11cm]{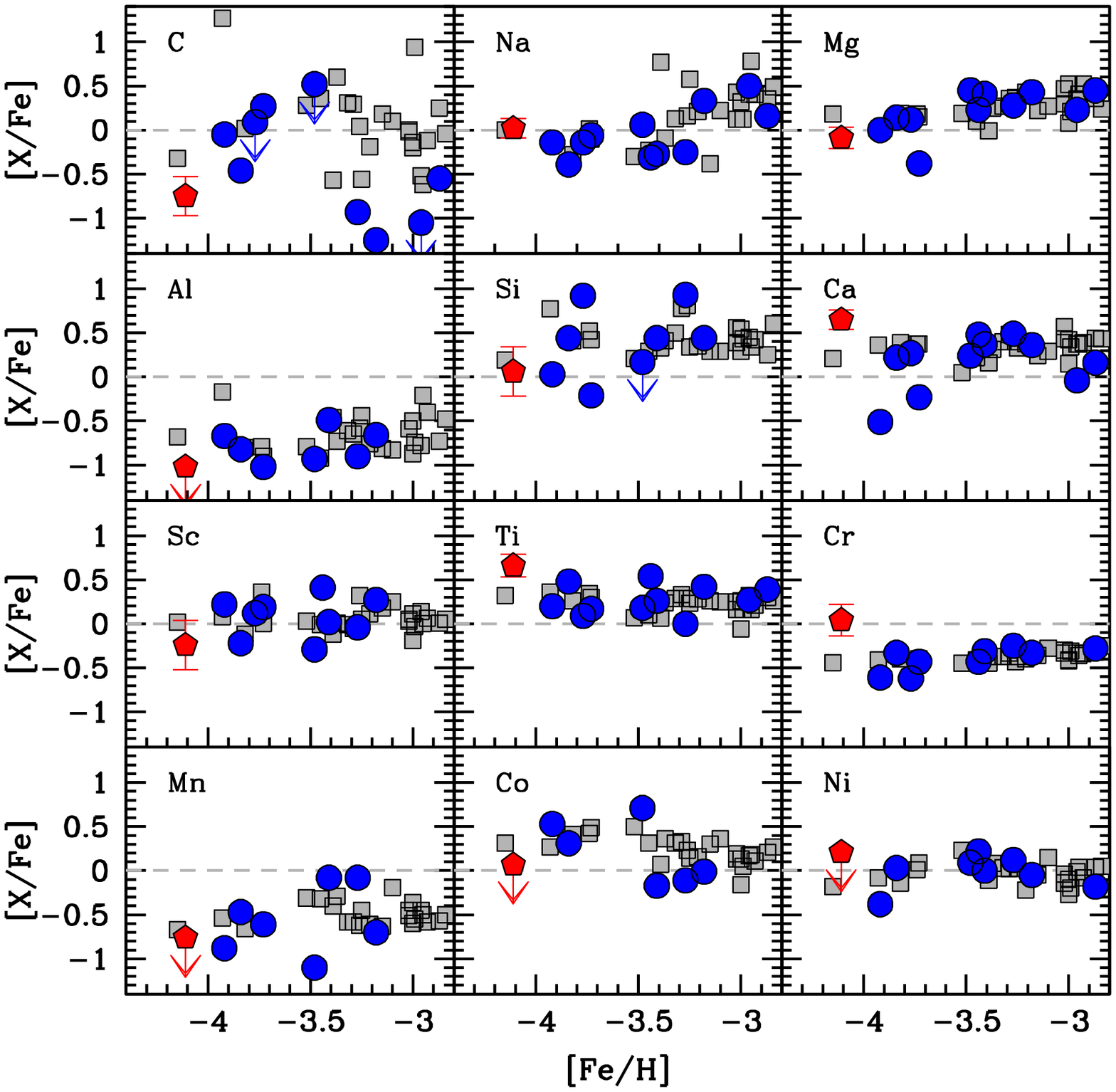} 
      \caption{Comparison of the LTE abundances of AS0039 (red pentagon) with literature values. Abundances for other Sculptor stars are shown with blue circles \citep{Starkenburg13,Jablonka15,Tafelmeyer10,Simon10,Frebel10}, while grey squares are stars in the Milky Way \citep{Cayrel04}. 
      }
      \end{figure}

\section{Additional tables}

Detailed information about the star AS0039 and the X-Shooter spectra used in this analysis are listed in Table~\ref{tab:obs}. List of the Fe lines used for the abundance analysis is given in Table~\ref{tab:linelist_fe}, and information for other elemental lines are in Table~\ref{tab:linelist}

\begin{table}
\caption{The observational log for the X-Shooter spectra of AS0039.}
\label{tab:obs}
\centering
\small
\begin{tabular}{l c}
\hline\hline
Star & AS0039 \\
\hline
RA  		&  00:58:45.64 \\
Dec.	& $-$33:42:24.4 \\
\hline
Obs. date		& 	2018-12-16 \\
Exp. time		&		2 x 1800\,s\\
Airmass (mean) &  1.1 \\
Seeing (mean) & 0.6''\\
\hline
Slit width (UVB) & 1.0'' \\
R$_{\text{UVB}}$		& 5\,400 \\
$\lambda_{\text{UVB}}$ & 3000-5600\,\AA \\
S/N$_{\text{UVB}}$ (4680\,\AA) & 38 pxl$^{-1}$\\
\hline
Slit width (VIS)  & 0.9'' \\
R$_{\textsl{VIS}}$	& 8\,900 \\
$\lambda_{\text{VIS}}$ & 5500-10200\,\AA \\
S/N$_{\text{VIS}}$ (6050\,\AA) & 42 pxl$^{-1}$\\
\hline
G 	& 16.932 $\pm$ 0.001  \\
BP 	& 17.525 $\pm$ 0.010  \\
RP  	& 16.073 $\pm$ 0.010 \\
\hline
$\mu_{\alpha^*}$ & $-0.01\pm0.06$ mas/yr\\
$\mu_\delta$  & $-0.16\pm0.05$ mas/yr\\
$v_{rad}$ & 135 $\pm$ 1 km/s\\
\hline
$T_\textsl{eff}$ & 4377 $\pm$ 81 K\\
$\log g $ & 0.8$\pm$ 0.1 \\
$v_{mic}$ & 2.0$\pm$ 0.1 km/s\\
\hline\hline
\end{tabular}
\end{table}

\begin{table}
\footnotesize
\begin{center}
\caption{Measurements of individual Fe lines.
}
\label{tab:linelist_fe}
\begin{tabular}{lrrrrr}
\hline
\noalign{\smallskip}
Species & \multicolumn{1}{c}{$\lambda / \mathrm{nm}$} & \multicolumn{1}{c}{$\chi_{\text{low}}$} & \multicolumn{1}{c}{$\log{g\,f}$} & \multicolumn{1}{c}{$\log\epsilon_{\text{LTE}}$} & \multicolumn{1}{c}{$\Delta_{\text{NLTE}}$ [dex]} \\
\noalign{\smallskip}
\hline
\noalign{\smallskip}
\ion{Fe}{1} & $ 381.584$ & $   1.490$ & $   0.237$ & \multirow{2}{*}{$    3.37$} & \multirow{2}{*}{$    0.17$} \\ 
\ion{Fe}{1} & $ 381.634$ & $   2.200$ & $  -1.196$ &  & \\ 
\noalign{\smallskip}
\hline
\noalign{\smallskip}
\ion{Fe}{1} & $ 382.043$ & $   0.860$ & $   0.119$ & \multirow{2}{*}{$    3.21$} & \multirow{2}{*}{$    0.19$} \\ 
\ion{Fe}{1} & $ 382.118$ & $   3.270$ & $   0.198$ &  & \\ 
\noalign{\smallskip}
\hline
\noalign{\smallskip}
\ion{Fe}{1} & $ 382.430$ & $   3.300$ & $  -0.043$ & \multirow{2}{*}{$    3.38$} & \multirow{2}{*}{$    0.18$} \\ 
\ion{Fe}{1} & $ 382.444$ & $   0.000$ & $  -1.362$ &  & \\ 
\noalign{\smallskip}
\hline
\noalign{\smallskip}
\ion{Fe}{1} & $ 382.588$ & $   0.920$ & $  -0.037$ & $    2.81$ & $    0.21$ \\ 
\ion{Fe}{1} & $ 382.782$ & $   1.560$ & $   0.062$ & $    2.70$ & $    0.24$ \\ 
\ion{Fe}{1} & $ 385.082$ & $   0.990$ & $  -1.734$ & $    3.00$ & $    0.17$ \\ 
\ion{Fe}{1} & $ 386.552$ & $   1.010$ & $  -0.982$ & $    3.75$ & $    0.07$ \\ 
\ion{Fe}{1} & $ 387.250$ & $   0.990$ & $  -0.928$ & $    3.47$ & $    0.12$ \\ 
\noalign{\smallskip}
\hline
\noalign{\smallskip}
\ion{Fe}{1} & $ 389.789$ & $   2.690$ & $  -0.736$ & \multirow{2}{*}{$    3.40$} & \multirow{2}{*}{$    0.18$} \\ 
\ion{Fe}{1} & $ 389.801$ & $   1.010$ & $  -2.018$ &  & \\ 
\noalign{\smallskip}
\hline
\noalign{\smallskip}
\ion{Fe}{1} & $ 390.648$ & $   0.110$ & $  -2.243$ & $    2.79$ & $    0.22$ \\ 
\ion{Fe}{1} & $ 392.026$ & $   0.120$ & $  -1.746$ & $    3.16$ & $    0.18$ \\ 
\ion{Fe}{1} & $ 392.291$ & $   0.050$ & $  -1.651$ & $    3.91$ & $    0.09$ \\ 
\ion{Fe}{1} & $ 400.524$ & $   1.560$ & $  -0.610$ & $    3.80$ & $    0.06$ \\ 
\ion{Fe}{1} & $ 404.581$ & $   1.490$ & $   0.280$ & $    2.83$ & $    0.20$ \\ 
\ion{Fe}{1} & $ 406.359$ & $   1.560$ & $   0.062$ & $    3.57$ & $    0.09$ \\ 
\ion{Fe}{1} & $ 407.174$ & $   1.610$ & $  -0.022$ & $    3.43$ & $    0.11$ \\ 
\ion{Fe}{1} & $ 413.206$ & $   1.610$ & $  -0.675$ & $    3.79$ & $    0.06$ \\ 
\noalign{\smallskip}
\hline
\noalign{\smallskip}
\ion{Fe}{1} & $ 414.342$ & $   3.050$ & $  -0.204$ & \multirow{2}{*}{$    3.63$} & \multirow{2}{*}{$    0.18$} \\ 
\ion{Fe}{1} & $ 414.387$ & $   1.560$ & $  -0.511$ &  & \\ 
\noalign{\smallskip}
\hline
\noalign{\smallskip}
\ion{Fe}{1} & $ 419.825$ & $   3.370$ & $  -0.457$ & \multirow{2}{*}{$    3.64$} & \multirow{2}{*}{$    0.20$} \\ 
\ion{Fe}{1} & $ 419.830$ & $   2.400$ & $  -0.719$ &  & \\ 
\noalign{\smallskip}
\hline
\noalign{\smallskip}
\ion{Fe}{1} & $ 420.203$ & $   1.490$ & $  -0.708$ & $    3.53$ & $    0.10$ \\ 
\ion{Fe}{1} & $ 423.594$ & $   2.420$ & $  -0.341$ & $    3.20$ & $    0.24$ \\ 
\ion{Fe}{1} & $ 425.012$ & $   2.470$ & $  -0.405$ & $    3.80$ & $    0.19$ \\ 
\ion{Fe}{1} & $ 425.079$ & $   1.560$ & $  -0.714$ & $    3.44$ & $    0.11$ \\ 
\ion{Fe}{1} & $ 426.047$ & $   2.400$ & $   0.109$ & $    3.56$ & $    0.20$ \\ 
\ion{Fe}{1} & $ 432.576$ & $   1.610$ & $   0.006$ & $    3.32$ & $    0.13$ \\ 
\ion{Fe}{1} & $ 438.355$ & $   1.490$ & $   0.200$ & $    3.05$ & $    0.17$ \\ 
\ion{Fe}{1} & $ 441.512$ & $   1.610$ & $  -0.615$ & $    2.73$ & $    0.26$ \\ 
\ion{Fe}{1} & $ 442.731$ & $   0.050$ & $  -2.924$ & $    3.44$ & $    0.17$ \\ 
\ion{Fe}{1} & $ 446.165$ & $   0.090$ & $  -3.210$ & $    3.37$ & $    0.18$ \\ 
\noalign{\smallskip}
\hline
\noalign{\smallskip}
\ion{Fe}{1} & $ 448.217$ & $   0.110$ & $  -3.501$ & \multirow{2}{*}{$    3.34$} & \multirow{2}{*}{$    0.19$} \\ 
\ion{Fe}{1} & $ 448.225$ & $   2.220$ & $  -1.482$ &  & \\ 
\noalign{\smallskip}
\hline
\noalign{\smallskip}
\ion{Fe}{1} & $ 452.861$ & $   2.180$ & $  -0.822$ & $    3.25$ & $    0.23$ \\ 
\noalign{\smallskip}
\hline
\noalign{\smallskip}
\end{tabular}
\end{center}
\end{table}

\begin{table}
\begin{center}
\footnotesize
\caption{Atomic data and measured abundances of individual lines.}
\label{tab:linelist}
\begin{tabular}{lrrrrr}
\hline
\noalign{\smallskip}
Species & \multicolumn{1}{c}{$\lambda / \mathrm{nm}$} & \multicolumn{1}{c}{$\chi_{\text{low}}$} & \multicolumn{1}{c}{$\log{g\,f}$} & \multicolumn{1}{c}{$\log\epsilon_{\text{LTE}}$} & \multicolumn{1}{c}{$\Delta_{\text{NLTE}}$ [dex]} \\
\noalign{\smallskip}
\hline
\noalign{\smallskip}
\ion{Li}{1} & $ 670.776$ & $   0.000$ & $  -0.002$ & \multirow{2}{*}{$    0.50$\tablenotemark{a}} & \multirow{2}{*}{$    0.03$} \\ 
\ion{Li}{1} & $ 670.792$ & $   0.000$ & $  -0.303$ &  & \\ 
\noalign{\smallskip}
\hline
\noalign{\smallskip}
\ion{Na}{1} & $ 588.995$ & $   0.000$ & $   0.117$ & $    2.24$ & $   -0.06$ \\ 
\ion{Na}{1} & $ 589.592$ & $   0.000$ & $  -0.194$ & $    2.01$ & $    0.06$ \\ 
\ion{Mg}{1} & $ 382.936$ & $   2.709$ & $  -0.231$ & $    3.59$ & $    0.06$ \\ 
\noalign{\smallskip}
\hline
\noalign{\smallskip}
\ion{Mg}{1} & $ 383.829$ & $   2.717$ & $  -1.530$ & \multirow{3}{*}{$    3.20$} & \multirow{3}{*}{$    0.12$} \\ 
\ion{Mg}{1} & $ 383.829$ & $   2.717$ & $   0.392$ &  & \\ 
\ion{Mg}{1} & $ 383.830$ & $   2.717$ & $  -0.355$ &  & \\ 
\noalign{\smallskip}
\hline
\noalign{\smallskip}
\ion{Mg}{1} & $ 517.268$ & $   2.712$ & $  -0.402$ & $    3.26$ & $    0.11$ \\ 
\ion{Mg}{1} & $ 518.360$ & $   2.717$ & $  -0.180$ & $    3.35$ & $    0.07$ \\ 
\ion{Al}{1} & $ 396.152$ & $   0.014$ & $  -0.323$ & $    1.30$\tablenotemark{a} & $    1.00$ \\ 
\ion{Si}{1} & $ 390.552$ & $   1.909$ & $  -1.041$ & $    3.46$ & $   -0.05$ \\ 
\ion{Ca}{1} & $ 422.673$ & $   0.000$ & $   0.244$ & $    2.59$ & $    0.07$ \\ 
\ion{Ca}{1} & $ 616.217$ & $   1.899$ & $  -0.090$ & $    2.82$ & $    0.15$ \\ 
\ion{Ca}{2} & $ 849.802$ & $   1.692$ & $  -1.416$ & $    3.01$ & $   -0.28$ \\ 
\ion{Ca}{2} & $ 854.209$ & $   1.700$ & $  -0.463$ & $    2.82$ & $   -0.07$ \\ 
\ion{Ca}{2} & $ 866.214$ & $   1.692$ & $  -0.723$ & $    2.94$ & $   -0.09$ \\ 
\ion{Sc}{2} & $ 424.682$ & $   0.315$ & $   0.242$ & $   -1.47$ & $    0.00$ \\ 
\ion{Sc}{2} & $ 431.408$ & $   0.618$ & $  -0.096$ & $   -0.95$ & $    0.00$ \\ 
\ion{Ti}{1} & $ 441.727$ & $   1.887$ & $  -0.020$ & $    1.64$\tablenotemark{b} & $    0.16$ \\ 
\ion{Ti}{1} & $ 447.124$ & $   1.734$ & $  -0.103$ & $    2.17$\tablenotemark{b} & $    0.16$ \\ 
\noalign{\smallskip}
\hline
\noalign{\smallskip}
\ion{Ti}{1} & $ 501.419$ & $   0.000$ & $  -1.220$ & \multirow{2}{*}{$    1.43$\tablenotemark{b}} & \multirow{2}{*}{$    0.57$} \\ 
\ion{Ti}{1} & $ 501.428$ & $   0.813$ & $   0.110$ &  & \\ 
\noalign{\smallskip}
\hline
\noalign{\smallskip}
\ion{Ti}{2} & $ 390.054$ & $   1.131$ & $  -0.200$ & $    1.51$ & $    0.36$ \\ 
\ion{Ti}{2} & $ 391.346$ & $   1.116$ & $  -0.420$ & $    1.43$ & $    0.34$ \\ 
\ion{Ti}{2} & $ 428.787$ & $   1.080$ & $  -1.790$ & $    1.74$ & $    0.17$ \\ 
\ion{Ti}{2} & $ 430.004$ & $   1.180$ & $  -0.440$ & $    1.38$ & $    0.33$ \\ 
\ion{Ti}{2} & $ 430.191$ & $   1.161$ & $  -1.150$ & $    1.62$ & $    0.24$ \\ 
\ion{Ti}{2} & $ 431.497$ & $   1.161$ & $  -1.100$ & $    2.04$ & $    0.25$ \\ 
\ion{Ti}{2} & $ 433.791$ & $   1.080$ & $  -0.960$ & $    1.44$ & $    0.19$ \\ 
\ion{Ti}{2} & $ 439.503$ & $   1.084$ & $  -0.540$ & $    1.58$ & $    0.24$ \\ 
\ion{Ti}{2} & $ 439.977$ & $   1.237$ & $  -1.190$ & $    1.73$ & $    0.26$ \\ 
\ion{Ti}{2} & $ 444.379$ & $   1.080$ & $  -0.720$ & $    1.09$ & $    0.02$ \\ 
\ion{Ti}{2} & $ 445.048$ & $   1.084$ & $  -1.520$ & $    1.83$ & $    0.17$ \\ 
\ion{Ti}{2} & $ 446.851$ & $   1.131$ & $  -0.600$ & $    1.37$ & $    0.02$ \\ 
\ion{Ti}{2} & $ 450.127$ & $   1.116$ & $  -0.770$ & $    1.26$ & $    0.24$ \\ 
\ion{Ti}{2} & $ 457.197$ & $   1.572$ & $  -0.320$ & $    1.07$ & $    0.12$ \\ 
\ion{Ti}{2} & $ 518.868$ & $   1.582$ & $  -1.050$ & $    1.68$ & $    0.15$ \\ 
\ion{Cr}{1} & $ 428.972$ & $   0.000$ & $  -0.361$ & $    1.35$ & $    0.89$ \\ 
\ion{Cr}{1} & $ 520.451$ & $   0.940$ & $  -0.208$ & $    1.90$ & $    0.64$ \\ 
\ion{Cr}{1} & $ 520.604$ & $   0.940$ & $   0.019$ & $    1.77$ & $    0.65$ \\ 
\ion{Cr}{1} & $ 520.843$ & $   0.940$ & $   0.158$ & $    1.17$ & $    0.66$ \\ 
\ion{Mn}{1} & $ 403.075$ & $   0.000$ & $  -0.470$ & $    0.70$\tablenotemark{a} & $    1.00$ \\ 
\ion{Mn}{1} & $ 403.306$ & $   0.000$ & $  -0.618$ & $    0.40$\tablenotemark{a} & $    1.00$ \\ 
\ion{Co}{1} & $ 399.530$ & $   0.923$ & $  -0.220$ & $    0.90$\tablenotemark{a} & $    1.00$ \\ 
\ion{Ni}{1} & $ 547.690$ & $   1.826$ & $  -0.890$ & $    2.30$\tablenotemark{a} & $    0.16$ \\ 
\ion{Sr}{2} & $ 407.771$ & $   0.000$ & $   0.167$ & $   -2.00$ & $    0.00$ \\ 
\ion{Sr}{2} & $ 421.552$ & $   0.000$ & $  -0.145$ & $   -2.38$ & $    0.00$ \\ 
\ion{Ba}{2} & $ 455.403$ & $   0.000$ & $   0.170$ & $   -3.20$\tablenotemark{a} & $    0.00$ \\ 
\noalign{\smallskip}
\hline
\noalign{\smallskip}
\end{tabular}
\end{center}
\tablenotetext{a}{upper limits are shown} 
\tablenotetext{b}{\ion{Ti}{1} excluded from the mean}
\end{table}

\end{document}